\documentclass[12pt,preprint]{aastex}

\usepackage{epsfig}

\slugcomment{To appear in ApJ}

\shorttitle{Detection of the 2175\AA\ Extinction Feature at $z=0.83$}
\shortauthors{Motta et al.}

\begin{document}

\title{Detection of the 2175\AA\ Extinction Feature at $z=0.83$}

{\author{V. Motta\altaffilmark{1,2}, E. Mediavilla\altaffilmark{1}, J. A. Mu\~noz\altaffilmark{1}, E. Falco\altaffilmark{3}, C.S. Kochanek\altaffilmark{3}, S. Arribas\altaffilmark{4,5}, B. Garc\'{\i}a-Lorenzo \altaffilmark{6}, A. Oscoz\altaffilmark{1}, M. Serra-Ricart\altaffilmark{1}}

\altaffiltext{1}{Instituto de Astrof\'{\i}sica de Canarias, V\'{\i}a L\'actea S/N, La Laguna 38200, Tenerife, Spain}
\altaffiltext{2}{Departamento de Astronom\'{\i}a, Facultad de Ciencias, Igu\'a 4225, Montevideo 11400, Uruguay}
\altaffiltext{3}{Harvard-Smithsonian Center for Astrophysics, 60 Garden Street, Cambridge, MA 02138, USA}
\altaffiltext{4}{Space Telescope Science Institute. 3700 San Martin Drive,
Baltimore, MD21218, USA. Affiliated with the Research and Science
Support Department of ESA. On leave from the Instituto de Astrof\'{\i}sica de
Canarias}
\altaffiltext{5}{Consejo Superior de Investigaciones Cient\'{\i}ficas}
\altaffiltext{6}{Isaac Newton Group of Telescopes, 38700 S/C de La Palma, Spain}

\begin{abstract}  
We determine the extinction curve in the $z_l=0.83$ lens galaxy of the
gravitational lens SBS0909+532 from the wavelength dependence of the
flux ratio between the lensed quasar images ($z_s=1.38$) from 3400
to 9200\AA.  It is the first measurement of an extinction curve at a
cosmological distance of comparable quality to those obtained within
the Galaxy.  The extinction curve has a strong 2175\AA\ feature,
a noteworthy fact because it has been weak or non-existent in most
estimates of extinction curves outside the Galaxy.  The extinction
curve is fitted well by a standard $R_V=2.1\pm0.9$ Galactic extinction
curve. If we assume standard Galactic extinction laws, the
estimated dust redshift of $z=0.88\pm0.02$ is in good agreement with
the spectroscopic redshift of the lens galaxy.  The widespread
assumption that SMC extinction curves are more appropriate models for
cosmological dust may be incorrect.
\end{abstract}


\section{INTRODUCTION}
Just as the properties of galaxies and stars evolve with cosmic time,
so must the properties of the interstellar medium.  In particular,
with changes in metallicity and ionizing backgrounds we might expect
the properties of dust to change with redshift.  Locally we can make
precise measurements of extinction curves as a probe of the dust
grains by comparing the spectral energy distributions (SED) of
reddened and unreddened stars of the same spectral type (the pair
method, \citet{massa83}), but the need to measure the spectra of
individual stars limits this approach to the Galaxy, its satellites
and its nearest neighbors.

The average Galactic extinction curve is well modeled over a broad
wavelength range as a function of only the ratio of total to selective
extinction, $R_V$ (e.g. Cardelli, Clayton \& Mathis 1989, hereafter
CCM; Fitzpatrick 1999).  These extinction curves include a spectral
feature, the 2175\AA\ bump, and appear to apply to a wide range of
interstellar dust environments in the Galaxy.  It is not, however, a
universally valid law.  While some lines of sight in the Magellanic
clouds show similar extinction laws, others, like the 30 Doradus
region of the Large Magellanic Cloud (LMC) show weaker 2175\AA\
features and steeper far UV extinction curves
\citep[e.g.,][]{fitzpatrick86}.  The most extreme deviations from the
Galactic extinction curves are found in the bar of the Small
Magellanic Cloud (SMC, Gordon \& Clayton 1998). Some newer studies
\citep[e.g.,][]{clayton00} have found lines of sight with weak
2175\AA\ features and steep far UV extinctions in the Galaxy as well.
M31 is the most distant galaxy to which the pair method can be
applied; the M31 extinction curve seems similar to that of the Galaxy
but may have a weaker 2175\AA\ bump \citep{bianchi96}.

The studies of extinction in other galaxies are in many cases based on
the examination of the attenuation induced by dust in the SED of the
galaxies.  But the physical properties of dust and its geometrical
distribution affect in a complex way the intrinsic SED of the galaxy
stellar population, which must be also modeled.  The extinction curves
inferred from these indirect methods are strongly model-dependent and,
in many cases, not unique due to model degeneracies
\citep{bianchi96,gordon97}.  In M33, \citet{gordon99} noticed that a
depression found in the SED should be related to the 2175\AA\ feature,
according to a radiative transfer model including attenuation by dust.
A very weak 2175\AA\ feature is also found by \citet{rosa94} by
comparing spectra of ionizing star clusters in HII regions of M101
with population synthesis models.  However, the modeling of the SED of
30 starburst galaxies \citep{gordon97} implies that the dust has
extinction properties similar to that of the SMC, lacking a 2175\AA\
feature.

Some efforts have also been made in AGNs to study the effects of
extinction by comparing the observed SEDs with conveniently selected
templates.  \citet{crenshaw01} have compared the spectra of the
Seyfert 1 galaxies NGC 3227 and NGC 4151 (assumed unreddened), and
found that the extinction in NGC 3227 increases very steeply toward
the far UV (steeper than in the SMC) and that there is no evidence of
the 2175\AA\ bump.  In QSOs, the comparison of the observed spectra
with composite spectra has not revealed traces of the 2175\AA\ feature
\citep{pitman00,vandenberk01}.  All the above-mentioned studies
suggest that the presence of a significant 2175\AA\ bump in the
average extinction curve of the Milky Way is rare \citep{pitman00}.
In any case, the validity of this or any other conclusion about the
behavior of the extinction beyond the Local Group is limited by the
lack of a direct method to obtain the extinction curve.  This
limitation has serious consequences for the study of galaxies
\citep[see][and references therein]{gordon00} and in cosmology
\citep[see][and references therein]{falco98}.

There are now two techniques which can extend the measurement of
precise extinction curves into the realm of cosmology.  One is based
on the use of the relationship between the variations in color and
absolute magnitude in SNIa light curves \citep{riess96}.  This
technique has been used to estimate several reddening ratios from
nearby supernovae which are in agreement with the MW average
extinction curve.  The other is an extension of the standard pair
method to obtain extinction curves in gravitational lens galaxies.
The method compares two different images of the same lensed QSO which
follow different paths through the lens galaxy \citep{nadeau91}.  As
in the pair method each image has the same intrinsic spectrum but
suffers from a different amount of extinction along each path through
the lens galaxy.  Using broad-band optical and NIR photometry,
\citet{falco99} measured differential extinctions in 23 gravitational
lens galaxies.  In spite of the coarse spectral sampling,
\citet{falco99} could fit the average Galactic extinction law to
several lens galaxies, finding values for $R_V$ in the range 0.64 to
7.2. This range includes some extreme values for $R_V$ whose physical 
significance is not obvious as the CCM law was inferred from values of $R_V$ 
in the 2.5 to 5.5 range.
\citet{falco99} also derived relatively accurate dust redshifts following
the technique proposed by \citet{jean98}.  These studies relied on
broad band photometry, which can lead to problems distinguishing
between extinction and gravitational microlensing and cannot provide
complete extinction curves over broad wavelength ranges.  These
problems can be overcome using spectra of the individual images,
particularly using integral field spectroscopy (IFS) to obtain a 2D
array of spectra of the lens and its images \citep{mediavilla98}.

In this article, we use IFS to search for the 2175\AA\ feature in a
cosmologically distant galaxy by selecting a gravitational lens where
the lens galaxy shows significant extinction and the redshifted
wavelength of the 2175\AA\ feature, $2175(1+z_l)$\AA\ is below the Lyman
break in the spectrum of the source quasar at $912(1+z_s)$\AA.  We
performed the experiment on the two-image, $1\farcs11$ separation
quasar lens SBS~0909+532 \citep{kochanek97}, where the $z_l=0.83$ lens
and $z_s=1.38$ source redshifts \citep{oscoz97,lubin00} satisfy the
wavelength criterion, and \citet{falco99} found there was significant
differential extinction $\Delta E(B-V) = 0.20 \pm 0.03$ based on the
image flux ratios measured in broad band imaging.

\section{DATA ANALYSIS}

\subsection{Observations and Data Reduction}

The spectra of SBS 0909+532 were taken on January 18, 2001 with the
2D-fiber system INTEGRAL \citep{arribas98}, at the 4.2-m William
Herschel Telescope (WHT).  INTEGRAL links the Nasmyth focus of the WHT
with the WYFFOS spectrograph.  We used the SB1 fiber bundle that
consists of a rectangular array of 175 fibers, each $0\farcs45$ in
diameter, covering an area of $7\farcs9\times 6\farcs7$ on the sky,
and a ring with diameter $90\arcsec$ formed by 30 fibers of the same
size as those in the inner array.  The distance between adjacent
fibers is $0\farcs5$ in the central rectangle.  At the entrance of the
spectrograph, the fibers are aligned to form a pseudo-slit.

We took three exposures of 1800s each. We used the 300lines/mm
grating, R300B, that yields a linear dispersion of
6.0\AA\,pixel$^{-1}$ covering the wavelength range from $3400$ to
$9200$\AA.  The seeing, measured from a guide star was $\sim
1\arcsec$.

For our data reduction, we used the {\tt IRAF}\footnote{{\tt IRAF} is
distributed by the National Optical Astronomy Observatories, which is
operated by the Association of Universities for Research in Astronomy,
Inc. (AURA) under cooperative agreement with the National Science
Foundation} software package.  The procedure included bias
subtraction, cosmic-ray rejection, spectra extraction, throughput
correction, wavelength calibration, and sky subtraction.  All these
steps were applied to each one of the three individual frames and to
an average frame obtained by combining them.  For additional details
on reduction and analysis of data taken with optical fibers, see
\citet{arribas91}.

In Figure~\ref{fig1} we show the 2D distribution of spectra from the
average frame over a spectral range from 6000 to 7200 \AA\ including
the MgII$\lambda$2798 emission line.  The positions of the two compact
components A and B are also shown.

\begin{figure}[h]
\plotone{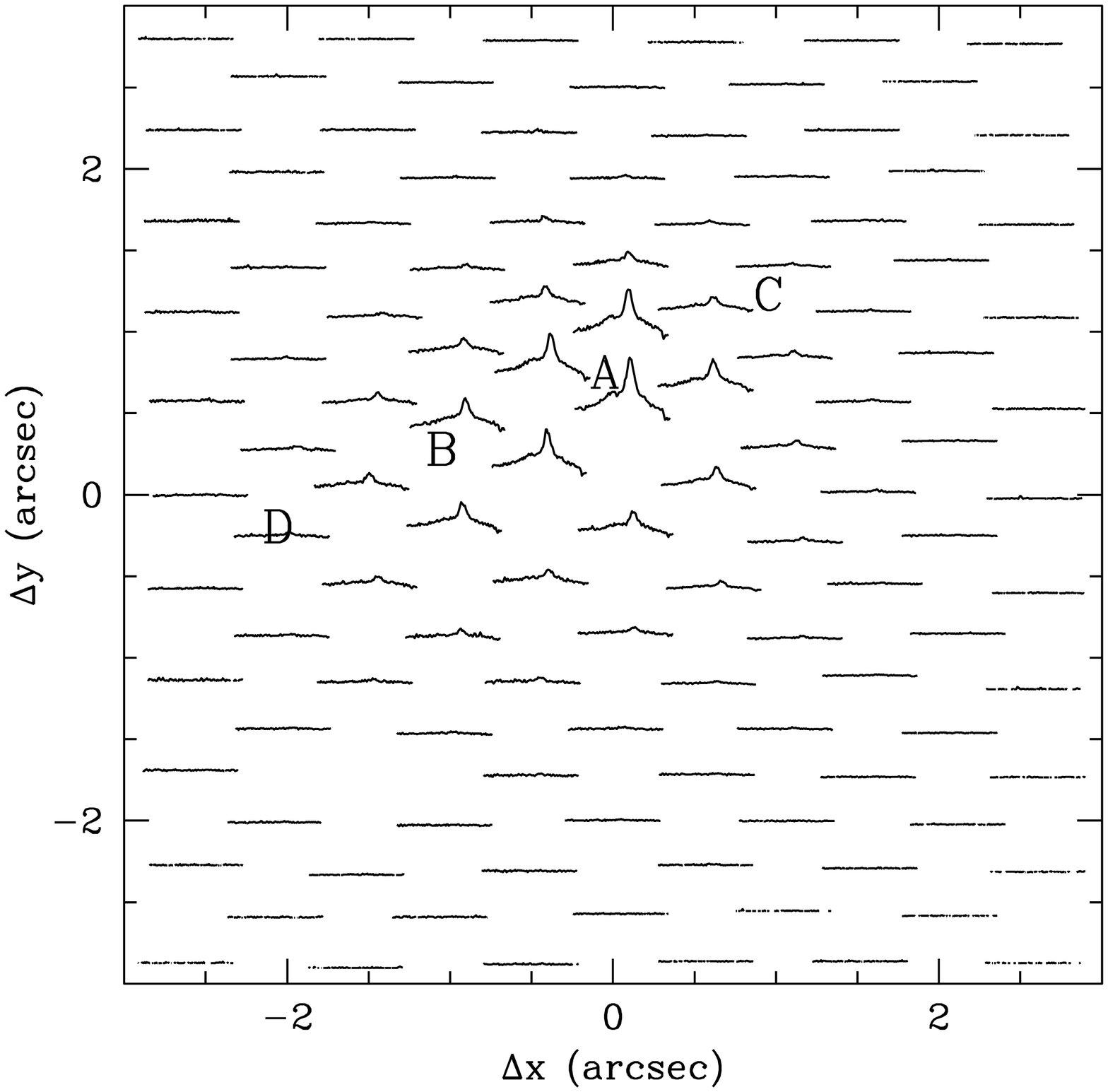}
\caption{The 2D distribution of spectra around the A and B components,
in the $6000$ - $7200$\AA\, wavelength range. The locations of the two
QSO images are marked A and B. C and D mark the locations where two
spectra were interpolated to correct the effects of seeing in the A
and B spectra (see section~\ref{interpolated})
\label{fig1}}
\end{figure}

\begin{figure}[h]
\epsfig{file=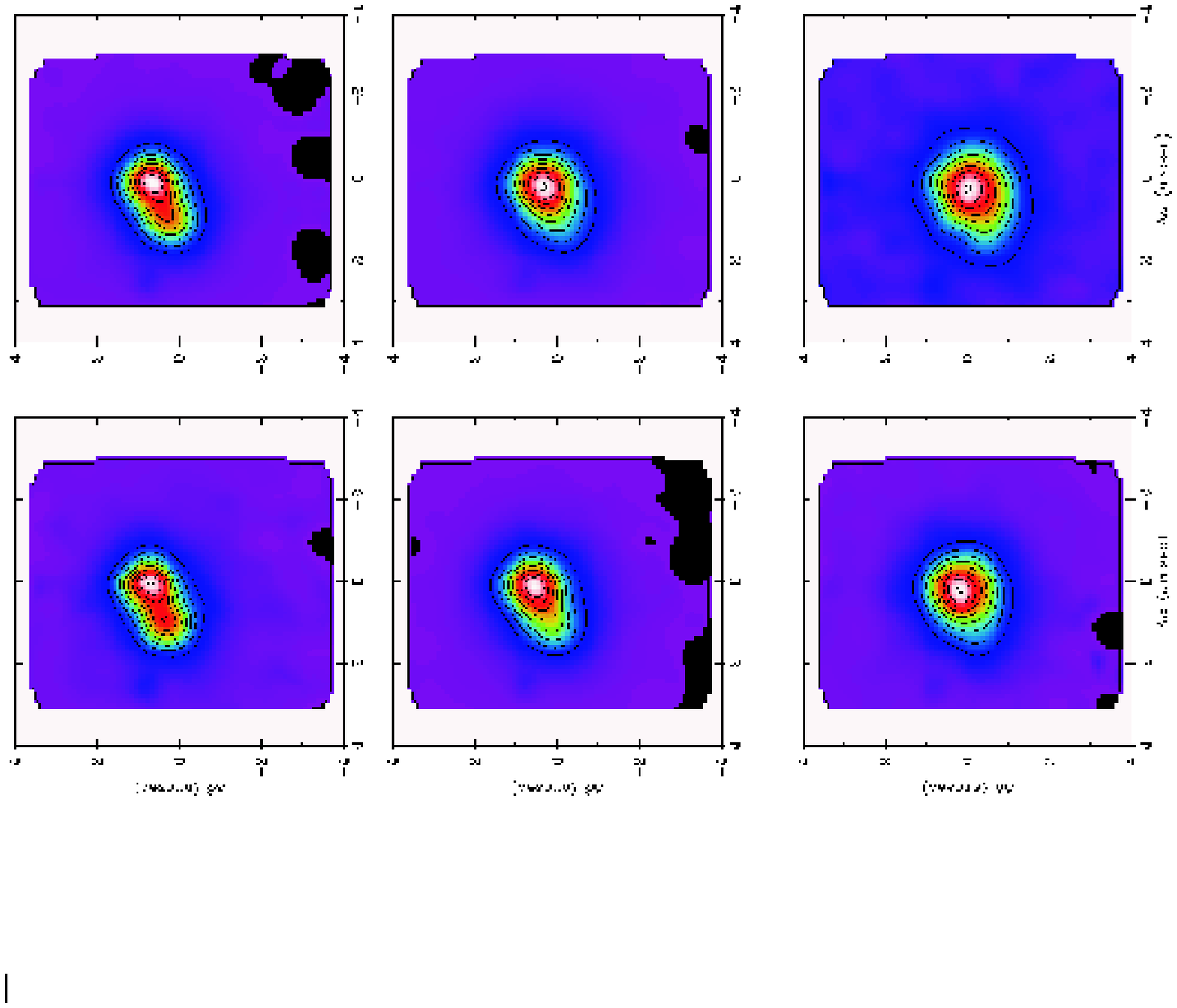,angle=270,width=13cm}
\caption{The plots shows 6 of the 20 continuum maps described in the
text. From top to bottom, left to right, the continuum central
wavelengths are 8727\AA, 7233\AA, 5739\AA, 4447\AA, 4006\AA, and
3487\AA.
\label{fig2}}
\end{figure}

\subsection{Continuum Maps}

From the 2D collection of spectra corresponding to the averaged frame
(a subset of them around components A and B is shown in
Figure~\ref{fig1}) we interpolated 20 continuum maps covering all the
observed spectral range.  The central wavelengths and widths of the
continua are included in Table~\ref{tab1}.  We fitted and removed the
strong emission lines present in the QSO spectra.

In Figure~\ref{fig2} we present six of these continuum maps
corresponding to the averaged frame centered at 8727\AA, 7233\AA,
5739\AA, 4447\AA, 4006\AA, and 3487\AA.  The two compact components of
SBS 0909+532 appear well separated in the reddest map (8727\AA,
seeing$\sim 1\farcs05$).  Notice the continuous fading of image B from
the red to the blue maps.  In fact, in the bluest map (3487\AA) the B
image is no longer visible.  This decrease in the B/A flux ratio
towards the blue is explained by dust absorption in the lensing galaxy
\citep{falco99}.

\begin{figure}[h]
\plotone{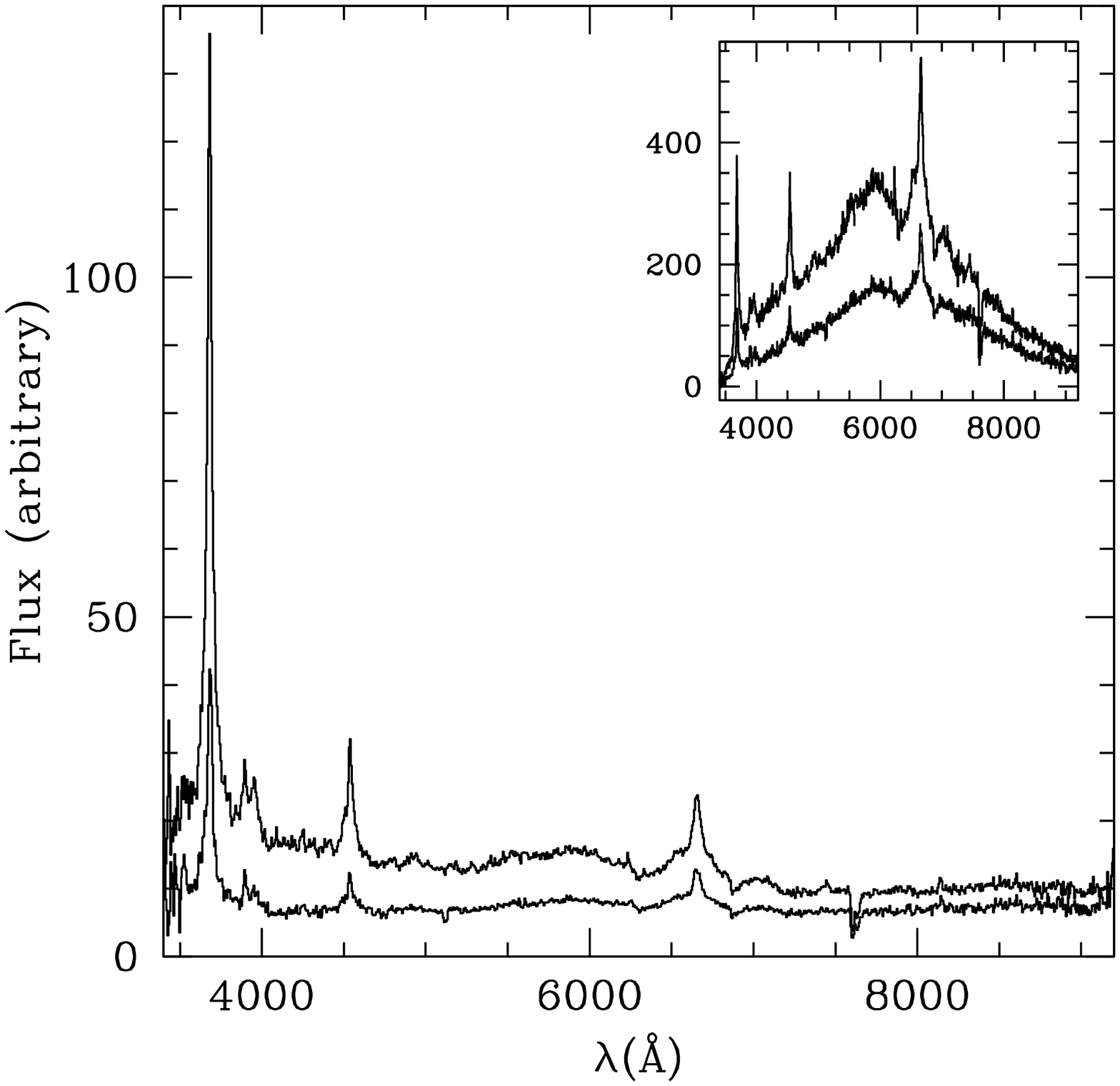}
\caption{Spectra corresponding to the A and B QSO images corrected for
the detector response. The small box shows the original spectra.
\label{fig3}}
\end{figure}

\subsection{A and B Spectra}

Because of differential atmospheric refraction (DAR), the location of
any of the components in the telescope focal plane changes with
wavelength.  In fact, we measured a displacement of about $0\farcs6$
(greater than the diameter of a fiber face) between the bluest and
reddest continuum maps.  Although the displacement is modest, it
implies that none of the spectra of Figure~\ref{fig1} can be exactly
identified with one of the components throughout the full spectral
range.  This is a general problem of spectroscopy in a large spectral
range that admits an {\it a posteriori} correction when 2D
spectroscopy is available \citep{arribas99}.  Following the procedure
described in this reference, we calculated a DAR model from the
location of component A in the 20 continuum maps distributed along the
full spectral range.  Next, from the original spectra we interpolated
a new 2D set of spectra corrected for DAR (see Arribas et al. 1999 for
the details).  From these spectra we obtained a map in the
3500-9000\AA\ spectral range.  We located the centroid of the A
component in this map and used the coordinates offset of B from A
given in \cite{lehar00} to infer the location of the B component.
Finally, we interpolated at these locations the A and B spectra (see
Figure~\ref{fig3}) from the ensemble of spectra corrected for DAR.
The changes from the uncorrected spectra are modest, as expected.

\section{RESULTS}

\subsection{Differential Extinction Curve from Continuum Images} 
\label{continuum}

To derive the B/A flux ratios from the ensemble of continuum maps
covering the full observed spectral range, we fit the maps using PSFs
derived from the continuum images of a star observed on the same
night.  We used as fitting parameters the separation between the two
QSO images, their relative intensity, the sky background, and the
blurring of the PSF due to the seeing.  We followed this procedure
with each one the three independent exposures.  After subtracting the
quasar model, the pattern of residuals show no emission from the
galaxy at any wavelength, so we conclude that our photometric model
with only two point-like sources reproduces successfully the observed
continua.  Acording to the HST data \citep{lehar00} the lens galaxy
is not detected in the $V$ band and its integrated magnitude is only 
5\% of component A flux in the $I$ band. The location of the two quasar 
components determined by our
photometric model is in excellent agreement with the results obtained
with the HST by \citet{lehar00} which strengthens the validity of our
photometric decomposition.

In Table 1 and Figure~\ref{fig4} we present the mean magnitude
differences, $m_B-m_A$, derived from the PSF fitting.  To estimate the
errors we have used the dispersion between the results obtained from
each independent exposure.  We include in this Figure the $m_B-m_A$
differences from \citet{lehar00} and \citet{kochanek97} corresponding
to broad band data; the agreement is excellent.

\begin{deluxetable}{cccc}
\tabletypesize{\scriptsize}
\tablecaption{\label{tab1}}
\tablewidth{0pt}
\tablehead{
\colhead{$\lambda$\tablenotemark{a}} & \colhead{$W$\tablenotemark{b}} & \colhead{$m_B$-$m_A$\tablenotemark{c}} & 
\colhead{$\Delta (m_B$-$m_A)$\tablenotemark{d}}} 
\startdata
3487& 252 &1.40 & 0.20 \\
3565& 96 &1.45 & 0.06 \\
3661& 96 &1.59 & 0.05 \\
3757& 96 &1.67 & 0.12 \\
3853& 96 &1.83 & 0.10 \\
3949& 96 &1.77 & 0.13 \\
4063& 96 &1.75 & 0.10 \\
4159& 96 &1.63 & 0.17 \\
4255& 96 &1.52 & 0.08 \\
4351& 96 &1.52 & 0.08 \\
4447& 96 &1.35 & 0.09 \\
4743& 498 &1.07 & 0.08 \\
5241& 498 &0.91 & 0.03 \\
5739& 498 &0.83 & 0.02 \\
6237& 498 &0.69 & 0.02 \\
6735& 498 &0.55 & 0.03 \\
7233& 498 &0.47 & 0.03 \\
7731& 498 &0.43 & 0.03 \\
8229& 498 &0.37 & 0.02 \\
8727& 498 &0.37 & 0.02 \\
\enddata
\tablenotetext{a}{Central wavelength of the continuum maps (\AA)}
\tablenotetext{b}{Wavelength bin of the continuum maps (\AA)}
\tablenotetext{c}{Average magnitude differences}
\tablenotetext{d}{Error in magnitude differences}
\end{deluxetable}

\begin{figure}[H] 
\plotone{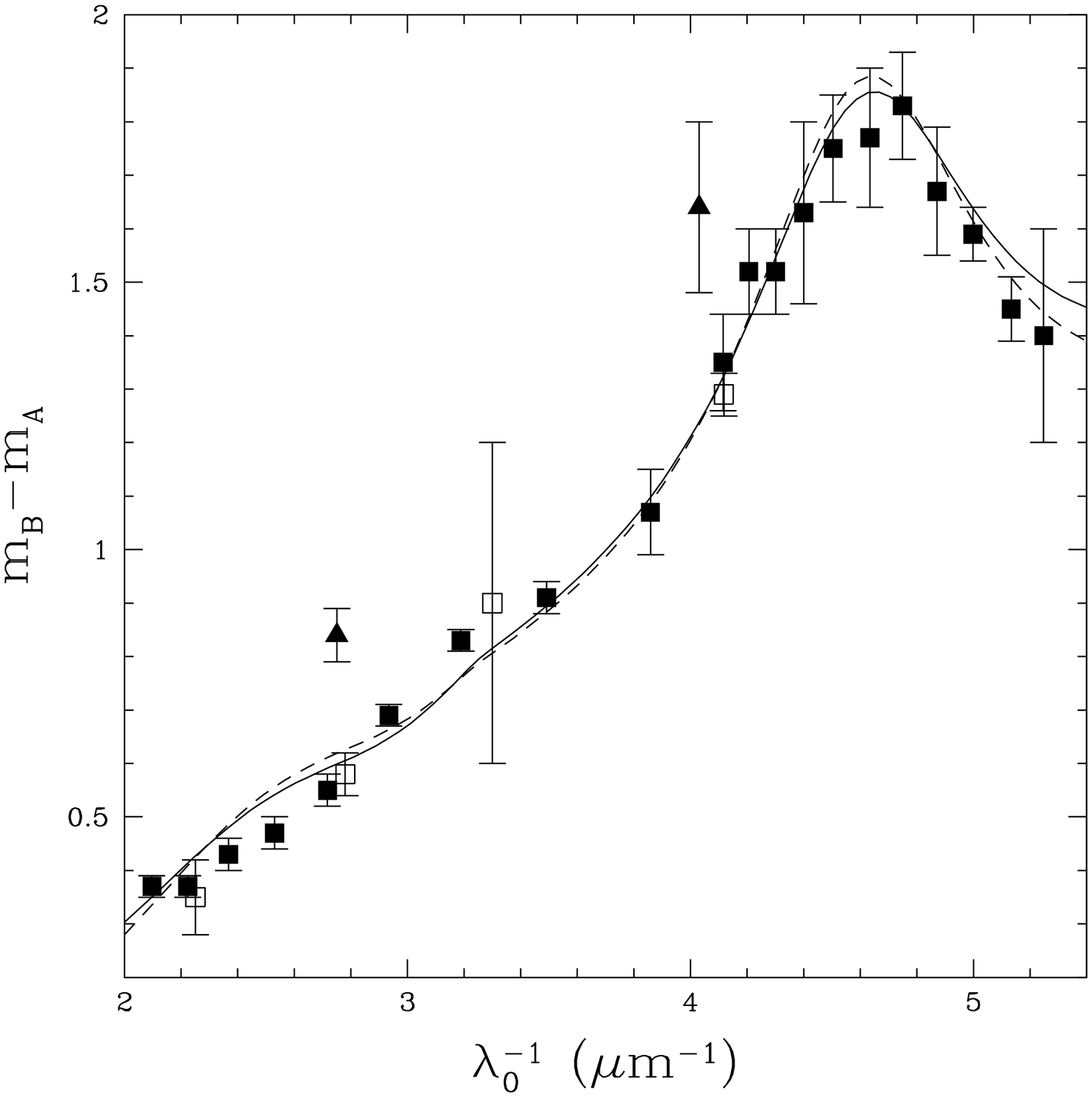} 
\caption{Magnitude differences $m_B-m_A$ obtained from continuum
images as a function of inverse wavelength in the lens galaxy rest
frame (filled squares). Open squares correspond to the $B, V, R, I$
broad-band observations \citep{kochanek97,lehar00}. The solid line is
the best fit of the analytical average MW extinction law to the
observed extinction curve (filled squares) for a fixed $z=0.83$. The
values $R_V=2.1\pm0.9$ and differential extinction
$E(B-V)=0.21\pm0.02$ were obtained in the fit. The filled triangles
correspond to the $m_B-m_A$ differences computed from the emission in
the MgII and CIII] lines. The dashed line corresponds to the best fit for 
$R_V=3.1$.
\label{fig4}} 
\end{figure}

\subsection{Differential Extinction Curve from Emission Line Images}

Proceeding in the same way as for the continua, we have also computed
the $m_B-m_A$ differences for the emission in the MgII$\lambda$2798
and CIII]$\lambda$1909 lines.  To estimate and subtract the continuum
below each line, we fitted a straight line to adjacent points.  
 The magnitude differences in the emission lines are approximately 
0.3 mag above that of the continuum ($m_B-m_A$, see Figure~\ref{fig4}). 
These differences are probably created by microlensing from the stars in the 
lens galaxy which can differentially magnify the continuum and emission line 
regions of the quasar because of their differing physical sizes \citep{schneider88}.  While there are too few emission lines to estimate an extinction curve, 
the emission lines are consistent with the extinction properties estimated 
from the continuum. 
We have also tried to compute the
$m_B-m_A$ difference from the CIV]$\lambda$1548, but the maps were too
noisy and the results of the PSF fitting were not consistent.

\subsection{Differential Extinction Curve from A and B Uncontaminated 
Spectra} \label{interpolated}

The method followed in Section~\ref{continuum} could be repeated with
narrower continua to obtain the differential extinction curve with
improved spectral sampling that could even match the spectral
resolution of the data.  However, the spectral sampling that can be
achieved with this method is, in practice, limited by the noise in
the data and the PSF.

To improve the spectral resolution we can try, alternatively, and only
because the lens galaxy is undetected, to obtain the extinction curve
directly from the $A$ and $B$ spectra after correcting them for the
cross-contamination induced by the seeing.  We can estimate the
contribution of $A$ to the spectrum measured at the position of $B$ by
measuring the spectrum at position $C$ located at the same distance
from $A$ as in the image $B$ but in the opposite direction (see
Figure~\ref{fig1}).  Similarly we obtain spectrum $D$ as an estimate
of the contribution to $A$ from $B$.  Our estimate of the
seeing-corrected spectra are the differences $A_u=A-D$ and $B_u=B-C$.

We can check the results by comparing the spectral ratio to the
estimates made by directly fitting the continuum maps.  They agree
reasonably well, with a $\sim 0.15$~mag drift between the blue and the
red wavelengths.  The drift appears to be due to small asymmetries in
the PSF. We can improve the agreement by defining
$A_u=A-(1-\epsilon)D$ and $B_u=B-(1+\epsilon)C$ with $\epsilon=0.1$ to
correct for the asymmetries.  Figure~\ref{fig5} shows our final
estimate of the magnitude differences
$[m_{B_u}-m_{A_u}]({1/\lambda_0})$ after smoothing by a
$\sigma=30$\AA\ Gaussian.  This curve is in good agreement with the
difference magnitudes obtained from the continuum images at all
wavelengths except in the vicinity of the emission lines.  As stated
in the previous section, this could arise from microlensing that is
not affecting the lines to the same degree as the continuum.  This
implies that, in some spectral regions (like the one including the
CIV$\lambda$1548 line in SBS 0909+532) observations made with
narrow-band filters could be problematic for obtaining the extinction
curves since the contribution from the continuum and the strong
emission lines must be separated.  (Notice also the presence of the Mg
II$\lambda\lambda$2796,2803 absorption doublet arising from the lens
galaxy).

\begin{figure}[h] 
\centerline{\epsfig{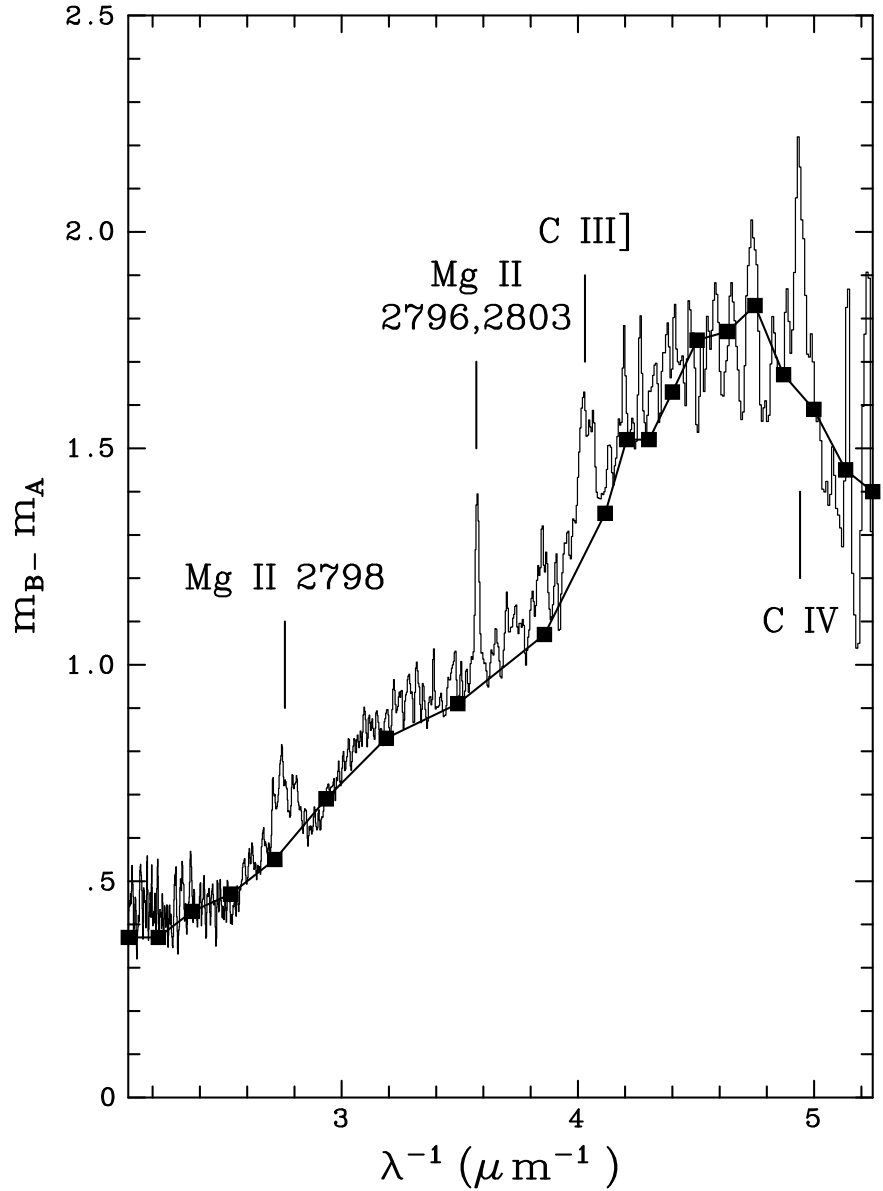}}
\caption{The continuous line is the $m_{B_u}-m_{A_u}$ magnitude
difference curve obtained from the A and B spectra. The abscissa is
the inverse wavelength at the lens galaxy rest frame. The squares
joined by straight lines correspond to the data presented in
Figure~\ref{fig4}. }
\label{fig5} 
\end{figure}

\section{DISCUSSION}

Assuming that any microlensing effects are achromatic, the magnitude
difference (Figure~\ref{fig4}) is a simple function of the
magnification ratio $\Delta M$ in magnitudes, the differential
extinction $\Delta E (B-V)$ and the extinction law $R(\lambda)$
\citep{falco99},
\begin{equation}
m_B(\lambda) -m_A(\lambda) = \Delta M + \Delta E(B-V) \, R \left(
\frac{\lambda}{1+z_l} \right).
\label{eq1}
\end{equation}

We fitted the $m_A-m_B$ data from the continua using equation
(\ref{eq1}) and the CCM parameterization of the extinction curve
$R(\lambda)$.  In a first step, the redshift is restricted to the
value determined directly from the absorption lines of the lens galaxy
($z=0.83$).  The best fit is obtained for $R_V=2.1\pm0.9$, 
 $\Delta M = -0.2\pm0.2$ and differential extinction 
$\Delta E(B-V)=0.21\pm0.02$.  A visual
inspection of the fit (see Figure~\ref{fig4}) shows reasonable
agreement with the CCM extinction curve for the MW ($\chi^2 / dof \sim
2$). In spite of the uncertainties in $R_V$, the fit of the
2175\AA\ bump is good in both shape and central wavelength.
When, in a second step, changes not only in $R_V$ but also in $z$
were allowed, we found that $R_V$ was very poorly determined, but we
obtained a useful estimate for the redshift, $z=0.88\pm 0.02$.  The
close agreement with the spectroscopic redshift of the lens galaxy
\citep[$z=0.83$,][]{oscoz97,lubin00} demonstrates the possibilities of
the dust-redshift technique \citep{jean98,falco99} and reinforces the
reliability of the extinction law.

The detection of a significant 2175\AA\ feature in the extinction
curve of a galaxy at $z=0.83$ is very noteworthy.  Many extragalactic
extinction studies have suggested that the 2175\AA\ feature is rare
\citep[see][and references therein]{pitman00}.  However, these studies
were based mainly on starburst galaxies and AGNs, where the dust is
subject to processing by radiation and shocks, similar to the
environments of star formation regions of the LMC or the SMC, where
the 2175\AA\ bump is weak.  Thus, the detection of this feature in the
extinction curve of the SBS 0909+532 lens (a normal early-type galaxy)
and its absence in active galaxies follows naturally under the
hypothesis of a connection between activity and suppression of the
2175\AA\ feature.

It has been proposed that the presence of the 2175\AA\ feature is
anti-correlated with the FUV extinction, with the feature weakening as
the FUV extinction steepens \citep{clayton00}.  Accordingly, the
rather low value $R_V=2.1$ (although not rare in the extragalactic
domain, see Falco et al. 1999 and references therein) that implies a
steep FUV rise would be inconsistent with the strong bump 
\citep[e.g. HD 210121,][]{clayton00}.  Within the uncertainty of the
determination of $R_V$ the average value for the MW ($R_V=3.1$) would
also be acceptable and in better accordance with the presence of a
significant bump.  Additional UV and IR data would be needed to better
answer this question.

\section{CONCLUSIONS}

The application of the standard pair method to spectroscopic
observations of the gravitationally lensed QSO SBS 0909+532 has
allowed us to obtain the first determination of an UV extinction law
beyond the Local Group.  Our conclusions are:

\begin{enumerate}
\item We have detected a significant 2175\AA\ bump in a $z=0.83$
galaxy.  So far, most studies of extinction in the extragalactic
domain have been based on starburst and AGN galaxies which seem to
exhibit SMC-like dust properties lacking a 2175\AA\ feature.  However,
the lens galaxy of SBS 0909+532 seems to contain dust like that of the
Galaxy.
\item The \citet{cardelli89} law provides a reasonable fit to the 
SBS\,0909+532 extinction curve with $R_V=2.1\pm0.9$.  This value includes,
at $\sim 1 \sigma$, the average MW law ($R_V=3.1$).  However, values of
$R_V \lesssim 2.1$ will be difficult to reconcile with the proposed
experimental correlation between weak bumps and steep FUV extinction.
\item The lens redshift estimated from the dust, $z_{dust}=0.88 \pm
0.02$, agrees well with the redshift inferred from the absorption
lines of the lens galaxy.
\end{enumerate}

We know of about 70 gravitational lenses, many of which show signs of
differential extinction \citep{falco99}.  By applying our techniques
to a wider sample of these lenses we should be able to provide a broad
outline of the properties and potentially the evolution of extinction
curves with cosmic time.

\acknowledgments The WHT is operated on La Palma by the Isaac Newton
Group (ING) of Telescopes in the Spanish Observatory Roque de los
Muchachos of the Instituto de Astrof\'{\i}sica de Canarias (IAC).  We
would like to thank the support astronomer A. Garc\'{\i}a P\'erez and
the night assistant N. Mahoney.  This work was supported by the P6/88
project of the IAC.  V. Motta acknowledges the support of a MUTIS
Fellowship from the Agencia Espa\~nola de Cooperaci\'on Internacional
(AECI).  E.E. Falco \& C.S. Kochanek were partially
supported by NASA through the grant GO-8252 from the Space Telescope
Science Institute, which is operated by the Association of
Universities for Research in Astronomy, Inc.

\clearpage

\end{document}